\documentstyle[aps]{revtex}
\begin{document}
\title{Relationships between two--particle overlap functions and the two--body
density matrix for many--fermion systems}
\author{A.N.Antonov$^{1}$, S.S.Dimitrova$^{1}$, M.V.Stoitsov$^{1}$, D.Van Neck$^{2}$%
and P.Jeleva$^{1}$}
\address{{\it $^{1}$\,Institute for Nuclear Research and Nuclear Energy, }\\
{\it Bulgarian Academy of Sciences, Sofia-1784, Bulgaria}\\
{\it $^{2}$\,Department of Subatomic and Radiation Physics,} \\
{\it University of Gent,} \\
{\it Proeftuinstraat 86, B--9000, Gent, Belgium}}
\maketitle

\begin{abstract}
Relationships are obtained connecting the two--nucleon overlap function of
the eigenstates in the $(A-2)$ particle system with the asymptotic behavior
of the two--body density matrix for the ground state of the A--particle
system.This makes it possible to calculate the two-body overlap functions,
spectroscopic factors and separation energies on the basis of a realistic
two-body density matrix. The procedure can be used in describing the $%
(e,e^{\prime }NN)$ and $(\gamma ,NN)$ reactions  where the two--body overlap
functions are a key ingredient in the analysis.

\medskip

\end{abstract}

\section{Introduction}

In this paper we examine the two--particle overlap functions in interacting
many--body systems and derive general relationships connecting them with the
ground state two--body density matrix. The procedure is based on the
asymptotic properties of the overlap functions in coordinate space, when the
distance between two of the particles and the center--of--mass of the
remaining ones becomes very large. This work can be considered as an
extension of the analysis presented in \cite{a3,a4,a6,a7,a1} where the
asymptotic behavior of the one--body density matrix and the single--particle
overlap functions was examined. Such an extension to the two-body sector is
of considerable interest in view of present--day experimental possibilities
in electromagnetically induced two--nucleon knock--out reactions like $%
(e,e^{\prime }NN)$ and $(\gamma ,NN)$.

The effects of ground--state NN correlations on two--nucleon knock--out
reactions have been intensively studied and discussed. This includes the
cases of the $(e,e^{\prime }pp)$ and $(e,e^{\prime }pn)$ reactions on $%
^{16}O $ and $^{12}C$ \cite{a2,a9,a10,a11,a12} and their contributions (for $%
^{12}C$ ) to the semi--exclusive $(e,e^{\prime }p)$ reaction \cite{a13}, as
well as of electroinduced \cite{c12} and of photonuclear processes \cite
{c13,c14,c15,a14,a15,a16,a16p}. It has been shown in \cite{a1,a9} that while
for $(e,e^{\prime }pn)$ reactions the central short-ranged correlations
(SRC) play only a marginal role, the two--proton knock--out in $(e,e^{\prime
}pp)$ processes exhibits a sizeable sensitivity to the ground--state
correlation effects. It was pointed out in \cite{a12} that the most
promising extraction of SRC effect shows up in the longitudinal structure
function which may be studied in the super--parallel kinematics. It was
found \cite{a11} that the knock--out of $^{1}S_{0}$ $pp$--pair dominates the
$^{16}O\,(e,e^{\prime }pp)$ reaction and this gives good perspectives for
extracting information on SRC in nuclei from such processes. It has been
shown also \cite{a13} that the two--hadron knock--out is a substantial
contribution to the $(e,e^{\prime })$ reaction mechanism above the
quasielastic peak. The correlation effects on $^{4}He\,(e,e^{\prime}d)\,d$
reaction cross--section have been studied in \cite{c12}. The important role
of the tensor correlations on the photonuclear cross--sections has been
pointed out in \cite{c13,c14,c15}. It was concluded in \cite{a15} that
combined analysis of $(\gamma ,p)$ and $(e,e^{\prime }p)$ reactions,
together with new data from $(\gamma ,NN)$ and $(e,e^{\prime }NN)$ processes
will lead to better understanding in electromagnetically induced knock--out
and that namely the $pp$--channel is more sensitive to the SRC \cite{a14,a16}%
.

The two--nucleon overlap functions and their properties are reviewed widely
e.g. in \cite{a17}. They have been used in \cite{a12} as one of the
components of the charge--current density, containing the information on the
nuclear structure in studies of two--nucleon knock--out and transitions to
the low--lying discrete final states of the residual nucleus. The overlap
functions used in \cite{a10} are constructed using phenomenological
single--particle wave functions and a correlation function from \cite{a18}.
As mentioned in \cite{a10} a more sophisticated treatment should be given in
principle on the basis of full calculations of the two--nucleon spectral
function where both long-- and short--range correlations are consistently
considered and hence spectroscopic factors are automatically included. In
\cite{a19} first calculations of the two--nucleon spectral function of $%
^{16}O$ have been performed. Long--range correlations are treated by a
Dressed RPA and SRC are included in the pair removal amplitude by adding
defect functions obtained from solutions of the Bethe--Goldstone equation
for the finite nucleus.

The aim of the present paper is to study to what extent observables
important for two--particle emission, such as the two--nucleon overlap
function, can be derived on the basis of the nuclear ground state two--body
density matrix. Our attempt is inspired by the possibility to extract
one--nucleon overlap functions from the one--body density matrix of the
nuclear ground state that has been proved in \cite{a3} and successfully
applied for calculating the overlap functions, spectroscopic factors and
separation energies associated with the bound states of the $(A-1)$%
---particle system \cite{a4,a7,a1}.


\section{Density matrices and overlap functions}

The one--and the two--body density matrices are defined in coordinate space
as:

\begin{equation}
\rho ^{(1)}(x,x^{\prime })=\langle {\Psi }^{(A)}|a^{+}(x)\,a(x^{\prime })|{%
\Psi }^{(A)}\rangle \,\,,  \label{eq1}
\end{equation}
and

\begin{equation}
\rho ^{(2)}(x_{1},x_{2};x_{1}^{\prime },x_{2}^{\prime })=\langle {\Psi }%
^{(A)}|a^{+}(x_{1})\,a^{+}(x_{2})\,a(x_{1}^{\prime })\,a(x_{2}^{\prime })|{%
\Psi }^{(A)}\rangle \,\,  \label{eq2}
\end{equation}
respectively, where $|{\Psi }^{(A)}\rangle $ is the antisymmetric $A$%
--fermion ground state normalized to unity and $a^{+}(x)$, $a(x)$ are
creation and annihilation operators at position $x$. The coordinate $x$
includes both the spatial coordinate ${\bf r}$ and the appropriate spin and
isospin coordinates. The matrices $\rho ^{(1)}$ and $\rho ^{(2)}$ are
trace--normalized to the number of particles and of pairs of particles:
\begin{equation}
{\it Tr}\,\rho ^{(1)}=\int \rho ^{(1)}(x)\,dx\;=\;A\,\,,  \label{norm1}
\end{equation}
\begin{equation}
{\it Tr}\,\rho ^{(2)}=\frac{1}{2}\,\int \rho
^{(2)}(x_{1},x_{2})\,dx_{1}\,dx_{2}\;=\;\frac{A(A-1)}{2}\,\,.  \label{norm2}
\end{equation}

Since $\rho ^{(1)}$ and $\rho ^{(2)}$ are hermitian matrices their
eigenstates $\psi_{\alpha }^{(i)}$ form a complete orthonormal set, in terms
of which $\rho^{(1)}$ and $\rho ^{(2)}$ can be decomposed as

\begin{equation}
\rho^{(1)} (x,x^{\prime })=\sum\limits_{\alpha =1}^{\infty }\lambda
_{\alpha}^{(1)}\psi _{\alpha}^{(1)*}(x)\psi^{(1)} _{\alpha}(x^{\prime })\;,\;
\label{nor1}
\end{equation}

\begin{equation}
\rho^{(2)} (x_1,x_2;x_1^{\prime },x_2^{\prime })=\sum\limits_{\alpha
=1}^{\infty }\lambda _{\alpha}^{(2)}\psi _{\alpha}^{(2)*}(x_1,x_2)\psi
^{(2)}_{\alpha}(x_1^{\prime},x_2^{\prime})\;.\;  \label{nor2}
\end{equation}
The eigenfunctions $\psi _{\alpha }^{(1)}$ and eigenvalues $\lambda _{\alpha
}^{(1)}$ are usually referred to as natural orbitals and natural occupation
numbers\cite{b1}. For fermionic systems the antisymmetry of the wave
functions imposes the constraint $0\leq \lambda _{\alpha }^{(1)}\leq 1$. In
analogy, the eigenfunctions $\psi _{\alpha }^{(2)}(x_{1},x_{2})$ are called
natural geminals and the associated real eigenvalues $\lambda _{\alpha
}^{(2)}$ -- natural geminal occupation numbers\cite{b2}. As a consequence of
the antisymmetry of the nuclear ground state, the $\lambda _{\alpha }^{(2)}$
obey the inequalities:
\begin{eqnarray}
0\;\leq \;\lambda _{i}^{(2)}\;\leq & (A-1)/2\;,\;\;\;\; & {\textstyle for}%
\;\;\;A\;\;{\textstyle odd}  \nonumber \\
0\;\leq \;\lambda _{i}^{(2)}\;\leq & A/2\;.\;\;\;\; & {\textstyle for}%
\;\;\;A\;\;{\textstyle even}  \label{eq3}
\end{eqnarray}

In uncorrelated (non--interacting) systems only the values $%
\lambda^{(i)}_{\alpha}\,=\,0,1$ appear for both the one-- and two--body
natural occupation numbers. Whereas the presence of correlations always
reduces the maximal one--body occupation number, this is not the case for
the two--body occupation numbers. The upper bound in (\ref{eq3}) is actually
only reached for systems which are in a sense maximally correlated, as e.g.
the occupation number of zero--coupled pairs in the seniority formalism in
the limit of large shell degeneracy.

Of more direct physical interest is the decomposition of the density
matrices in terms of the overlap functions between the $A$ particle ground
state and the eigenstates of the $(A-1)$ and $(A-2)$ particle systems, since
overlap functions can be probed in exclusive knock--out reactions.

Inserting a complete set of $(A-1)$ eigenstates $|\alpha (A-1)\rangle $ into
eq.(\ref{eq1}) one gets
\begin{equation}
\rho ^{(1)}(x,x^{\prime })=\sum\limits_{\alpha }\varphi _{\alpha
}^{(1)*}(x)\varphi _{\alpha }^{(1)}(x^{\prime })\;,\;  \label{obdm}
\end{equation}
where $\varphi _{\alpha }^{(1)}(x)=\langle \alpha (A-1)|a(x)|\Psi (A)\rangle$
is the one-body overlap function associated with the state $|\alpha
(A-1)\rangle $. The spectroscopic factors are then defined by the norm
\begin{equation}
S_{\alpha }^{(1)}=\langle \varphi _{\alpha }|\varphi _{\alpha }\rangle .
\label{specf1}
\end{equation}

The two--nucleon overlap functions are defined as the overlap between the
ground state of the target nucleus ${\Psi }^{(A)}$ and a specific state ${%
\Psi }_{\alpha }^{(C)}$ of the residual nucleus ($C=A-2$) \cite{a17}:
\begin{equation}
\Phi _{\alpha }(x_{1},x_{2})=\langle {\Psi }_{\alpha
}^{(C)}|a(x_{1})\,a(x_{2})|{\Psi }^{(A)}\rangle \,\,.  \label{ovf2}
\end{equation}
The two--particle spectroscopic factor is analogously defined by the norm
\begin{equation}
S_{\alpha }^{(2)}=\langle \Phi _{\alpha }|\Phi _{\alpha }\rangle
\label{spectf2}
\end{equation}
and the two--body density matrix reads
\begin{equation}  \label{tbdm}
\rho ^{(2)}(x_{1},x_{2};x^{\prime}_{1},x^{\prime}_{2}) =\sum\limits_{\alpha
}\Phi _{\alpha }^{*}(x_{1},x_{2})\Phi _{\alpha
}(x^{\prime}_{1},x^{\prime}_{2}) \, . \\
\end{equation}
As in the case of the single--particle spectroscopic factors where $%
S_{\alpha }^{(1)}\,\leq \,\lambda _{max}^{(1)}$ \cite{a3}, one can find that
$S_{\alpha }^{(2)}\,\leq \,\lambda _{max}^{(2)}$. Therefore, both one- and
two-particle spectroscopic factors cannot exceed the corresponding maximal
natural occupation numbers.


\section{Relationships between overlap functions and density matrices}

It has been shown in \cite{a3} that the one-body overlap functions
associated with the bound states of the $(A-1)$ system can be expressed in
terms of the ground state one-body density matrix of the $A$ nucleon system.
For example, in the case of a target nucleus with $J^{\pi }=0^{+}$, the
lowest $(n_{0}lj)$ bound state overlap function is determined by the
asymptotic behavior $(a\rightarrow \infty )$ of the corresponding partial
radial contribution ${\rho _{lj}(r,r}^{\prime }{)}$ of the one-body density
matrix:
\begin{equation}
\varphi _{n_{0}lj}(r)={\frac{{\rho _{lj}(r,a)}}{{C_{n_{0}lj}~\exp
(-k_{n_{0}lj}\,a})/a}}~,  \label{ovf0}
\end{equation}
where the constants ${C_{n_{0}lj}}$ and ${k_{n_{0}lj}}$ are completely
determined by ${\rho _{lj}(r,r}^{\prime }{)}$. In this way, both $\varphi
_{n_{0}lj}(r)$ and ${k_{n_{0}lj}}$ define the separation energy
\begin{equation}
\epsilon _{n_{0}lj}\equiv E_{n_{0}lj}^{(A-1)}~-~E_{0}^{(A)}=\frac{\hbar
^{2}~k_{n_{0}lj}^{2}}{2m}~  \label{sepe1}
\end{equation}
and the spectroscopic factor $S_{n_{0}lj}=\langle \varphi _{n_{0}lj}\mid
\varphi _{n_{0}lj}\rangle $. The procedure gives also the next bound state
overlap functions with the same multipolarity if they exist and its
applicability has been demonstrated in Refs. \cite{a4,a6,a7,a1}.

Similar procedure but applied to the two-body density matrix is of
significant physical interest for analyzing properties of transfer reactions
when two nucleons are removed from the target's ground state ${\Psi }^{(A)}$
leaving the residual $C=A-2$ system in a state ${\Psi }_{\alpha }^{(A-2)}$.
The procedure is possible again due to the particular asymptotic properties
of the two-body overlap functions (\ref{ovf2}). They satisfy the general
equation \cite{a17}:
\begin{equation}
\displaystyle{\ }\left( -{\frac{\hbar ^{2}}{{2m}}}\nabla _{1}^{2}-{\frac{%
\hbar ^{2}}{{2m}}}\nabla _{2}^{2}+{\it v}_{12}+\epsilon _{\alpha
}^{(2)}\right) \Phi _{\alpha }({\bf r}_{1},{\bf r}_{2})=\sigma ({\bf r}_{1},%
{\bf r}_{2}{)}\,\,,  \label{tbgeq}
\end{equation}
where ${\epsilon _{\alpha }^{(2)}\,=\,E_{\alpha }^{(A-2)}\,-\,E_{0}^{(A)}}$,
${\it v}_{12}$ is the internal di--nucleon interaction and $\sigma ({\bf r}%
_{1},{\bf r}_{2})$ is the nonlocal residual source term which contains the
interaction between the two extra nucleons and the C--nucleus.

The condition to obtain the asymptotic solution of eq.(\ref{tbgeq}) for the
two--body overlap function are considered in \cite{a17,a21,a22}. In \cite
{a21} the restrictions on the two--body interactions in the case of the
system of three nonrelativistic spinless neutral particles have been
formulated and the coordinate asymptotics of the discrete spectrum wave
functions have been studied. In \cite{a17,a22} the asymptotic behavior of
wave functions and overlap functions have been investigated generalizing the
amplitudes of Merkuriev \cite{a21} and different types of overlap functions
necessary in calculations of direct transfer amplitudes have been shown. In
case when a cluster of two like nucleons (neutrons or protons) unbound to
the rest of the system is transferred with a simultaneous transfer of both
nucleons, as e.g., in reactions $^{18}O-2n\,\rightarrow \,^{16}O$, or $%
^{16}O-2p\,\rightarrow \,^{14}C$, the following hyperspherical type of
asymptotics is valid for the two-body overlap functions \cite{a17,a21,a22}:
\begin{equation}
\Phi (r,R)\longrightarrow N\,e{xp}\left\{ -\sqrt{\frac{4m|E|}{\hbar ^{2}}%
\left( R^{2}+\frac{1}{4}r^{2}\right) }\right\} \,\left( R^{2}+\frac{1}{4}%
r^{2}\right) ^{-5/2}\;,  \label{ov2as}
\end{equation}
where ${\bf r}={\bf r}_{1}-{\bf r}_{2}$ and ${\bf R}=({\bf r}_{1}+{\bf r}%
_{2})/2$ are the magnitudes of the relative and center of mass coordinates,
respectively, $m$ is the nucleon mass and $E$ is the two-nucleon separation
energy
\begin{equation}
E\,=\,E^{(A)}-E^{(C)}\,=\,-E_{C}^{A}.  \label{ov2as1}
\end{equation}

In the case of a target nucleus with $J^{\pi }=0^{+}$ the two--body overlap
function (\ref{ovf2}) can be written in the form:
\begin{equation}
\Phi _{AJM}^{C}(x_{1},x_{2})=\sum_{LS}\left\{ \Phi _{\alpha JLS}({\bf r}_{1},%
{\bf r}_{2}){\otimes }\chi _{S}(\sigma _{1},\sigma _{2})\right\} _{JM}\,\,,
\label{tenzexp}
\end{equation}
where
\begin{eqnarray}
\chi _{SM_{S}}(\sigma _{1},\sigma _{2}) &=&\left\{ \chi _{\frac{1}{2}%
}(\sigma _{1})\otimes \chi _{\frac{1}{2}}(\sigma _{2})\right\} _{SM_{S}}
\nonumber \\
&=&\sum\limits_{m_{s_{1}}m_{s_{2}}}\left( \frac{1}{2}\,m_{s_{1}}\,\frac{1}{2}%
\,m_{s_{2}}|SM_{S}\right) \chi _{\frac{1}{2}m_{s_{1}}}(\sigma _{1})\,\chi _{%
\frac{1}{2}m_{s_{2}}}(\sigma _{2})  \label{tenzexp1}
\end{eqnarray}
and $\Phi _{\alpha JLSM_{L}}({\bf r}_{1},{\bf r}_{2})$ is the spatially
dependent part of the overlap function. As suggested in \cite{a17} it is
possible to perform a decomposition into angular momenta ${\bf l}={\bf l}_{r}
$ and ${\bf L}_{R}$ (${\bf L}={\bf l}+{\bf L}_{R}$), corresponding to the
relative and center of mass coordinates:
\begin{equation}
\Phi _{\alpha JSLM_{L}}({\bf r},{\bf R})=\sum_{lL_{R}}\Phi _{\alpha
JSLlL_{R}}(r,R)\left\{ Y_{L_{R}}(\widehat{{\bf R}})\otimes Y_{l}(\widehat{%
{\bf r}})\right\} _{LM_{L}}\;.  \label{LL}
\end{equation}
Then the two--body density matrix has the form:
\begin{eqnarray}
\rho ^{(2)}(x_{1},x_{2};x_{1}^{\prime },x_{2}^{\prime }) &=&\sum_{JM}%
\mathop{\sum_{LS}}_{L^{\prime }S^{\prime }}\mathop{\sum_{lL_{R}}}_{l^{\prime
}L_{R}^{\prime }}\mathop{\rho^{(2)}_{JSLlL_{R}}}_{\;\;\;\;\;\;S^{\prime
}L^{\prime }l^{\prime }L_{R}^{\prime }}(r,R;r^{\prime },R^{\prime })\,\times
\nonumber \\
&&A_{SLlL_{R}}^{JM\,\,\,*}(\sigma _{1},\sigma _{2};\widehat{{\bf r}},%
\widehat{{\bf R}})\,A_{S^{\prime }L^{\prime }l^{\prime }L_{R}^{\prime
}}^{JM}(\sigma _{1}^{\prime },\sigma _{2}^{\prime };\widehat{{\bf r}^{\prime
}},\widehat{{\bf R}^{\prime }})\,\,,  \label{tddec}
\end{eqnarray}
where the radial part of the density matrix is
\begin{equation}
\mathop{\rho^{(2)}_{ JSLlL_{R}}}_{\;\;\;\;\;\;S^{\prime }L^{\prime
}l^{\prime }L_{R}^{\prime }}(r,R;r^{\prime },R^{\prime })=\sum_{\alpha }\Phi
_{\alpha JSLlL_{R}}^{*}(r,R)\,\Phi _{\alpha JS^{\prime }L^{\prime }l^{\prime
}L_{R}^{\prime }}(r^{\prime },R^{\prime })  \label{ro21}
\end{equation}
and the spin--angular function is:
\begin{equation}
A_{SLlL_{R}}^{JM}(\sigma _{1},\sigma _{2};\widehat{{\bf r}},\widehat{{\bf R}}%
)=\left\{ \left\{ Y_{L_{R}}(\widehat{{\bf R}})\otimes Y_{l}(\widehat{{\bf r}}%
)\right\} _{L}\otimes \chi _{S}(\sigma _{1},\sigma _{2})\right\} _{JM}.
\label{tdso}
\end{equation}
In eqs. (\ref{tenzexp})--(\ref{ro21}) $\alpha $ is the number of the state
of the residual nucleus with a given total momentum $J$.

Let us further consider the diagonal part of eq.(\ref{ro21}):
\begin{equation}
\rho _{JSL{l}L_{R}}^{(2)}(r,R;r^{\prime },R^{\prime })= \sum_{\alpha }\Phi
_{\alpha JSLlL_{R}}^{*}(r,R)\, \Phi _{\alpha JSLlL_{R}}(r^{\prime
},R^{\prime }).  \label{ro21d}
\end{equation}
Eq.(\ref{ov2as}) implies that for large $r^{\prime}=a$ and $R^{\prime}=b$ a
single term $\alpha_0$ (corresponding to the smallest two-nucleon separation
energy) will dominate in the sum of the right hand side of eq.(\ref{ro21d}%
). As a consequence the two-body overlap function $\Phi _{\alpha_0
JSLlL_{R}}(r,R)$ can be expressed in terms of the two-body density matrix:
\begin{eqnarray}
\Phi _{\alpha_0 JSLlL_{R}}(r,R)\,&=&\,\frac{\rho _{JSLlL_{R}}^{(2)}(r,R;a,b)%
}{\Phi _{\alpha_0 JSLlL_{R}}(a,b)} \\
&=&\,\frac{\rho _{JSLlL_{R}}^{(2)}(r,R;a,b)}{N\,e{xp}\left\{ -k\sqrt{\left(
b^{2}+\frac{1}{4}a^{2}\right) }\right\} \,\left( b^{2}+\frac{1}{4}%
a^{2}\right) ^{-5/2}}.  \label{main}
\end{eqnarray}
Here the asymptotic form (\ref{ov2as}) of the two-body overlap function is
used with a constant $k$ which defines the two-nucleon separation energy (%
\ref{ov2as1}).

Therefore, if the two-body density matrix $\rho ^{(2)}$ is known, the
constant $k$ entering eq.(\ref{main}) can be determined by the slope of the
partial radial contribution $\rho _{JSLlL_{R}}^{(2)}(r,R;a,b)$ at large $a$
and $b$, while the unknown coefficients $N$ can be obtained from the
asymptotic behavior of the spatially diagonal part $\rho
_{JSLlL_{R}}^{(2)}(a,b;a,b)$. In this way, the constant $k$ will give
information about the two-nucleon separation (\ref{ov2as1}), and the norm of
the overlap function (\ref{main}) about the associated two--particle
spectroscopic factor. The relationships obtained can be useful in analyzing
the processes such as $(e,e^{\prime }NN)$ and $(\gamma ,NN)$ which are
studied widely nowadays. Of course, the method will be reliable when
realistic density matrices are considered.

In conclusion, the derivation of eq.(\ref{main}) makes it possible to find
a solution of the complicated problem about the relationship between the
two-body overlap functions and the two-body density matrix. The use of the
asymptotics of the overlap function in the particular case of two like
nucleons unbound to the rest of the system (\ref{ov2as}) can help in
analyzing the overlap functions for two--nucleon knock--out processes on the
basis of correlated two-body density matrix for the target nucleus. In our
future work, which is now in progress, the density matrices obtained in the
Jastrow--type model \cite{a20,a5} are used for actual calculations of the
two--particle overlap functions.

The authors thank Dr. C.Giusti for the helpful discussion and the Bulgarian
National Science Foundation for the partial financial support [contract $%
\Phi $-- 527].

\end{document}